\documentclass[aps,prd,twocolumn,superscriptaddress]{revtex4}
\usepackage{color}
\usepackage{graphicx}
\usepackage[hypertex]{hyperref}

\def\beq{\begin{equation}}
\def\eeq{\end{equation}}
\def\bea{\begin{eqnarray}}
\def\eea{\end{eqnarray}}

\def\msun{M_{\odot}}
\def\bwt{\begin{widetext}}
\def\ewt{\end{widetext}}
\usepackage{amssym,aas_macros}

\begin{document}

\title{Stellar Black Holes and the Origin of Cosmic Acceleration}

\author{Chanda Prescod-Weinstein}\email{cweinstein@perimeterinstitute.ca}
\affiliation{Perimeter Institute
for Theoretical Physics, 31 Caroline St. N., Waterloo, ON, N2L 2Y5, Canada}
\affiliation{Department of Physics and Astronomy, University of Waterloo, Waterloo, ON, N2L 3G1, Canada}
\author{Niayesh Afshordi}\email{nafshordi@perimeterinstitute.ca}
\affiliation{Perimeter Institute
for Theoretical Physics, 31 Caroline St. N., Waterloo, ON, N2L 2Y5, Canada}
\author{Michael L. Balogh}\email{mbalogh@uwaterloo.ca}
\affiliation{Department of Physics and Astronomy, University of Waterloo, Waterloo, ON, N2L 3G1, Canada}

\date{\today}
\preprint{astro-ph/yymmnnn}

\begin{abstract}
The discovery of cosmic acceleration has presented a unique challenge for cosmologists. As observational cosmology forges ahead, theorists have struggled to make sense of a standard model that requires extreme fine tuning. This challenge is known as the cosmological constant problem. The theory of gravitational aether is an alternative to general relativity that does not suffer from this fine-tuning problem, as it decouples the quantum field theory vacuum from geometry, while remaining consistent with other tests of gravity. In this paper, we study static black hole solutions in this theory and show that it manifests a UV-IR coupling: Aether couples the spacetime metric close to the black hole horizon, to metric at infinity. We then show that using the {\it Trans-Planckian ansatz} (as a quantum gravity effect) close to the black hole horizon, leads to an accelerating cosmological solution, far from the horizon. Interestingly, this acceleration matches current observations \textcolor{black}{for  stellar mass black holes}. Based on our current understanding of the black hole accretion history in the Universe, we then make a prediction for how the effective dark energy density should evolve with redshift, which can be tested with future dark energy probes.

\end{abstract}
\maketitle

\section{Introduction}
\label{sec:intro}

The discovery of recent acceleration of cosmic expansion was one of the most surprising findings in modern cosmology \cite{Riess:1998cb, Perlmutter:1998np}. The standard cosmological model (also known as the concordance model) drives this expansion with a cosmological constant (CC). While the CC is consistent with (nearly) all current cosmological observations \footnote{See \cite{2008arXiv0812.2244A} for an account of observational anomalies in the standard cosmological model.}, it requires an extreme fine-tuning of more than 60 orders of magnitude, known as {\it the cosmological constant problem} \cite{1989RvMP...61....1W}.

\textcolor{black}{In the context of the concordance cosmological model, there are (at least) three different aspects of the CC problem. For decades, physicists worried about why the value of the cosmological constant/vacuum energy seemed to be nearly zero by particle physics standards (known as the {\it old CC problem})\cite{Straumann:2002tv}, and the conventional wisdom was that it should vanish exactly, as a result of a yet unknown symmetry of nature. The accelerated cosmic expansion has thus challenged us to address this question on two new fronts. First is the {\it new CC problem}: why is the vacuum energy density so close to zero, but non-vanishing?  Second is the {\it coincidence problem}: Why did the dark energy dominance and structure formation happen at approximately coincident times?}

The race is on to simultaneously address these three questions. A popular alternative approach to the cosmological constant is a model that modifies Einstein's theory of gravity. Traditionally, this involves adding higher order curvature terms to the geometric side of Einstein's equation. However, in \cite{Afshordi:2008a}, one of us proposed a novel approach to modified gravity. This model introduces {\it gravitational aether}, as a necessary ingredient to decouple the quantum field theory vacuum from gravity while simultaneously satisfying other tests of gravity. Unlike many models of modified gravity, the gravitational aether model modifies the energy-momentum content of the spacetime, instead of adding higher order curvature terms.

In this model, the right hand side of the Einstein field equation is modified as:
\beq
(8\pi G')^{-1} G_{\mu\nu} = T_{\mu\nu} -\frac{1}{4}T^{\alpha}_\alpha g_{\mu\nu} + p'(u'_\mu u'_\nu +g_{\mu\nu}),\label{eq::aether}
\eeq
where $G'$ is $4/3$ times the Newton's constant, and $p'$ and $u'_{\mu}$ are aether pressure and four-velocity that are fixed by requiring the conservation of the energy-momentum tensor, $T_{\mu\nu}$, and the Bianchi identity. As argued in \cite{Afshordi:2008a}, while consistent with precision tests of gravity, this theory is preferred to the standard model by the combination of cosmological observations (with the notable exception of $^4$He primordial abundance).

In this paper, we pursue a detailed understanding of static spherical black hole solutions in the gravitational aether theory. The solution we find is, at first glance, a perturbed Schwarzschild metric. However, upon closer inspection we find that this perturbation is divergent both near to and far away from the horizon (where we refer to an infinite redshift surface as a horizon). Thus the static solution in the presence of gravitational aether is fundamentally different from Schwarzschild, which can be characterized as a UV-IR connection: the metric near and far from the horizon is set by the same integration constant. Here, we will explore possible meanings of this property, and whether the cosmological behavior is set by a {\it Trans-Planckian} ansatz close to the black hole horizon.

We note that the static black hole solution found here also applies to the cuscuton models \cite{Afshordi:2006ad,Afshordi:2007yx} which have the same energy-momentum tensor as the gravitational aether in the limit of vanishing cuscuton potential.

In Sec. \ref{sec:BH_metric}, we introduce our gravitational aether black hole solution. We describe the properties of the solution, including a preferred coordinate system and the location of the event horizon. We also establish asymptotic properties of the black hole, which are characterized by the same integration constant both close in and far away from the horizon of the black hole.

Sec. \ref{trans-planckian} explores the Trans-Planckian ansatz, as a way to fix the aforementioned integration constant, through quantum gravity effects close to the horizon. We suggest a way to connect the presence of black holes to the existence of a pervasive pressure that behaves like dark energy on cosmological scales.

In Sec. \ref{sec:Global}, we present a study of the contribution that many such black holes would make to the global/cosmological structure of space-time, while Sec. \ref{sec::history} provides a census of average black hole mass through cosmic history, which translates into a prediction for the history of cosmic acceleration.

Finally, we will discuss open questions and future prospects in Sec. \ref{sec:conclusions}.

Throughout the paper, we use the natural Planck units: $\hbar = c = G_{N} = k_B = 1$. Moreover, we will replace pressure $p'$ by $3p/4$ in Eq. (\ref{eq::aether}), so that the vacuum field equations for the aether theory resembles general relativity sourced by a perfect fluid with pressure $p$ and zero density.

\section{Black Hole in Gravitational Aether}
\label{sec:BH_metric}

We find a solution for the static black hole in the Gravitational Aether model using assumptions similar to those that lead to the Schwarzschild solution. Namely, we assume a spacetime with no matter content, and we assume spherical symmetry. Given that the aether takes fluid form, the metric in this model is the same as the general static, spherically symmetric metric that describes the interior of a star, as modeled by a perfect fluid. The only notable divergence from the star model is the absence of a matter density, leaving an energy-momentum tensor of the following form:
\begin{equation}
  T_{\mu\nu} = p(u_\mu u_\nu + g_{\mu\nu}).
  \label{e1}
\end{equation}
We find the following metric
\begin{equation}
        ds^2 = -e^{2\phi} dt^{2} + \left(1-\frac{2m}{r}\right)^{-1} dr^{2} + r^{2} d\Omega^{2}.
\label{e2}
\end{equation}
With components obeying the following differential equation, known as the Tolman-Oppenheimer-Volkoff equations \cite{2004sgig.book.....C}:
\bea
\frac{d\phi}{dr} = \frac{m+4\pi r^3 p}{r(r-2m)},
\label{e3}
\\
\frac{dp}{dr} = \frac{-p(m+4\pi r^3 p)}{r(r-2m)}.
\label{e3b}
\eea
We see immediately that $\exp(\phi)$ and $p$ are inversely related:
\begin{equation}
p=p_0 e^{-\phi},
\label{hydro}
\end{equation}
where $p_0$ is an integration constant. Notice that Eq. (\ref{hydro}) is equivalent to the condition of hydrostatic equilibrium for aether, and is valid independent of the assumption of spherical symmetry, for any static spacetime \footnote{\textcolor{black}{This follows from the relativistic Euler equation:$ (\rho+p){\bf u\cdot\nabla u} = -\nabla_{\perp} p$, assuming a static spacetime and zero density, $\rho =0$.}}. Now, we may rewrite Eqn. \ref{e3}:
\begin{equation}
\frac{d\phi}{dr} = \frac{m+4\pi r^3 p_0 e^{-\phi}}{r(r-2m)}
\label{e5}
\end{equation}

We can solve this equation by noting that it is a first-order inhomogeneous linear differential equation in $e^{\phi}$, with the standard solution:
\begin{equation}
e^{\phi(r)} = 4\pi p_0\left(1-\frac{2m}{r}\right)^{\frac{1}{2}} \left[\int \frac{(1-\frac{2m}{r})^{-1/2} r^2}{r-2m} dr + {\rm const.}\right].
\label{e7}
\end{equation}

\begin{figure}
\includegraphics[width=\linewidth]{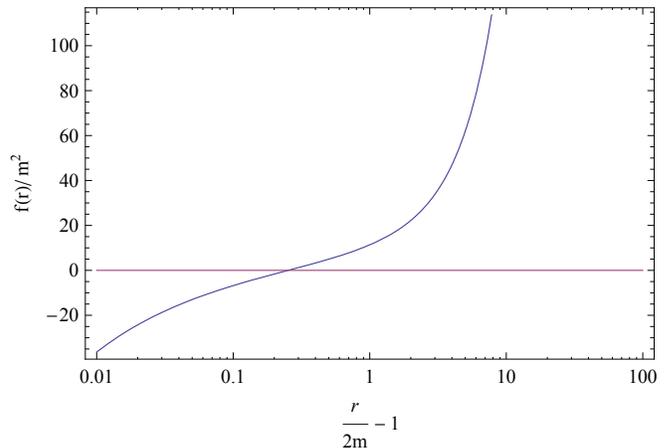}
 \caption
{\textcolor{black}{Function $f(r)$ [Eq. (\ref{e9})] as a function of the distance from the Schwarzschild radius ($=2m$). The deviation from the Schwarzschild metric is proportional to $p_0
f(r)$, where $p_0$ is the integration constant. If $p_0$ is
small, as we argue in Section \ref{trans-planckian}, the corrections only become important
at the horizon, and on cosmological scales. }}\label{fig-fr}
\end{figure}

To put this into more familiar terms, we can set the constant, so that we recover the Schwarzschild solution as $p_0 \rightarrow 0$:
\begin{equation}
e^{\phi(r)} = \left(1-\frac{2m}{r}\right)^{\frac{1}{2}}\left[4\pi p_0 f(r) + 1\right],
\label{e8}
\end{equation}
where $f(r)$ is given by:
\bea
f(r)=\frac{1}{2}\left(1-\frac{2 m}{r}\right)^{-1/2} \left(-30 m^2+5 m r+r^2\right)\nonumber\\+\frac{15}{2} m^2  \ln\left[\frac{r}{m}-1+\frac{r}{m}\left(1-\frac{2m}{r}\right)^{1/2}\right],
\label{e9}
\eea
and is shown in Fig. (\ref{fig-fr}). In the limit where $r$ is large ($r \gg m$):
\begin{equation}
f(r)=\frac{r^2}{2}+3mr+{\cal O}[m^2].
\label{e10}
\end{equation}
while close to the ``Schwarzschild horizon'' we find:
\begin{equation}\label{e11}
f(r)=-8\frac {\sqrt{2} m^{5/2}}{\sqrt{-2 m+r}}+{\cal O}[m^{3/2}(r-2 m)^{1/2}].
\end{equation}

Thus the correction to the Schwarzschild metric dominates in both UV and IR regimes (corresponding to close to and far from the BH horizon). This a nice tie, even for arbitrarily small values of the integration constant $p_0$. Therefore, a very suggestive conclusion is that, unlike in general relativity, the gravitational aether ties the formation of black hole horizons to cosmological dynamics.

 But then, is there really an event horizon for this spacetime? Looking at the trace of the Einstein's equation, we find that the Ricci scalar is proportional to the pressure of aether, $p$, which is in turn inversely proportional to the $00$ component of the metric, $e^{\phi}$. We define the surface where $e^{\phi} \rightarrow 0$ as the black hole horizon. Therefore the pressure at the horizon, and thus the Ricci scalar, goes to infinity ($p \propto {\cal R} \rightarrow \infty$) implying that this surface coincides with a real metric singularity (as opposed to a coordinate singularity).

 It appears that we have established that any static event horizon in a theory of gravitational aether (like the one we have modeled) coincides with a real metric singularity. In a traditional formulation of general relativity, such a scenario may be given to ambiguous physical interpretation. Cognizant of the fact that a modified gravity will display properties divergent from traditional relativity, we expect that such a picture is best contextualized by a more comprehensive theory of quantum gravity.

 {\it Indeed, any process (for example, quantum gravity) that alleviates/regulates metric singularities, will inevitably remove event horizons from the theory of gravitational aether.} In other words, static event horizons cannot exist in a UV completion of gravitational aether. This is independent of the assumption of spherical symmetry, and only relies on the aether hydrostatic equilibrium condition (\ref{hydro}). \textcolor{black}{However, we note that, as the singularity is a null surface, the spacetime does not violate the {\it weak cosmic censorship principle}.}

Back to the spherical aether black hole spacetime (\ref{e8}), we now
notice that the static metric solution is only well-defined for $r \geq
2m$, as the solution becomes complex inside the Schwarzschild radius,
$r<2m$. More surprisingly, for negative values of $p_0$, unlike a Schwarzschild black hole, a free-falling observer can reach this boundary within a finite {\it coordinate} time. The reason is that the redshift of a static source at the Schwarzschild radius is now finite as seen by distant observers \footnote{Here, distant observers are located at $2m \ll r \ll (-p_0)^{-1/2}$.}:
\bea
1+z &=& e^{-\phi} \simeq \left[\left(1-\frac{2m}{r}\right)^{1/2}-32\pi p_0 m^2\right]^{-1}\nonumber\\ &<& 1+z_{\rm max} = -\frac{1}{32\pi p_0 m^2}.\label{zmax}
\eea

As to what happens inside $r=2m$, it is clear that our current choice of coordinates do not give us a physical metric for $r< 2m$. However, is it possible that with an appropriate choice of coordinate, we can analytically continue the static solution beyond the Schwarzschild radius? Indeed, we can define a new radial coordinate:
\bea
&\lambda &\equiv \int_{2m}^r dr' \sqrt{g_{rr}} = \int_{2m}^r dr' \left(1-\frac{2m}{r'}\right)^{-1/2} \nonumber\\&= &2\left[2 m (r-2m)\right]^{1/2}+{\cal O}\left[(r-2m)^{3/2} m ^{-1/2}\right],
\eea
which is equivalent to the constant-time proper radial distance. In terms of $\lambda$, the metric takes the form:
\beq
ds^2 = -e^{2\phi} dt^2 + d\lambda ^2 + r(\lambda)^2 d\Omega^2,
\eeq
where
\bea
e^{\phi} = -32\pi p_0 m^2 + \frac{\lambda}{4m} + {\cal O}[p_0 \lambda^2, \lambda^3 m^{-3/2}], \\
r(\lambda) = 2m + \frac{\lambda^2}{8m} + {\cal O}[\lambda^4/m^2].
\eea
In other words, the metric is analytic and real in terms of the new radial coordinate, $\lambda$, at and beyond the Schwarzschild radius, which now corresponds to $\lambda =0$. Moreover, a static event horizon, which as we argued corresponds to a real curvature singularity, now exists for all (small) values of $p_0$, as $e^{\phi} =0$ at:
\beq
\lambda_{H} \simeq 128 \pi p_0 m^3.
\eeq

\textcolor{black}{In the next section, we study the implications of this solution for cosmology. However, we shall postpone a full investigation of the causal structure of this spacetime, as well as its possible analytic continuations, to future studies.}

\section{Trans-Planckian Ansatz and Cosmic Acceleration}\label{trans-planckian}

In the last section, we saw that within spherical spacetimes in the gravitational aether theory, the integration constant $p_0$ ties the geometry close to the horizon to the geometry at infinity. While, in the classical theory, $p_0$ is an arbitrary integration constant, here we speculate that its value is fixed by quantum gravity effects, especially since the horizon is now a curvature singularity, where quantum gravity effects should become important.

We first note that the temperature of sources that fall through the Schwarzschild horizon, as seen by distant observers, approaches the Hawking temperature \cite{NouriZonoz:1998td}:
\beq
T_{H} = \frac{1}{8\pi m},
\eeq
Furthermore, we assume that the maximum rest-frame temperature of sources is comparable to the Planck temperature (or one in Planck units):
\beq
T_{\rm max} = \theta_{P} = {\cal O}[1].
\eeq
Here, $\theta_P$ is a dimensionless constant that measures $T_{\rm max}$ in units of Planck temperature, which we shall call the Trans-Planckian parameter.
We then adopt the {\it Trans-Planckian ansatz}, which is the idea that the maximum redshift at Schwarzschild radius (Eq. \ref{zmax}) is roughly set by the ratio of the Planck to Hawking temperatures:
\beq
1+z_{\rm max} = -\frac{1}{32\pi p_0 m^2} = \frac{T_{\rm max}}{T_{H}} = 8\pi \theta_{P} m,
\eeq
or
\begin{figure}
\includegraphics[width=\linewidth]{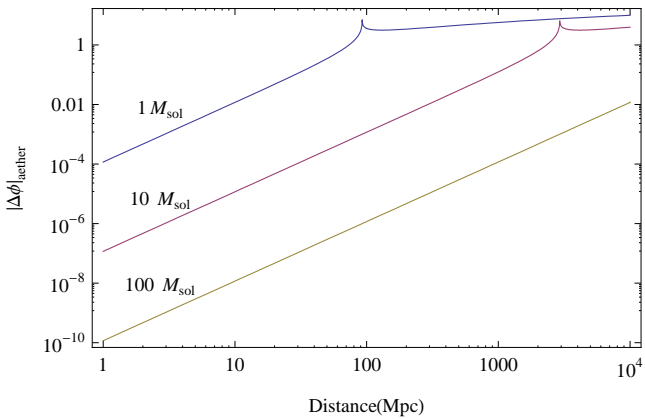}
 \caption
{\textcolor{black}{Predicted large distance deviation from the vacuum Schwarzschild solution for 1, 10, and 100 $\msun$ black holes, based on the Trans-Planckian ansatz. Here, we assumed $\theta_P = 100$ in Eq. (\ref{p0_m_eq}) for non-rotating black holes to find $p_0$, which is then plugged into Eq. (\ref{e8}) to find the metric. As pointed out in the text, the corrections become important on today's cosmological horizon scale for solar/stellar-mass black holes.} }\label{fig-dphi}
\end{figure}
\beq
p_0 = -\frac{1}{256\pi^2 \theta_{P} m^3}.\label{p0_m_eq}
\eeq
With this ansatz, we further see that:
\beq
\lambda_H = -\frac{1}{2\pi \theta_{P}} = {\cal O}[1],
\eeq
i.e. the event horizon is roughly a Planck length away from the Schwarzschild radius. Equivalently, the short-distance aether corrections to the Schwarzschild metric only become important at about a planck distance from the horizon/singularity, which is a reasonable expectation from a possible quantum gravitational mechanism.

While this may imply that tests of strong gravity close to the horizon of a black hole may have a hard time testing the influence of aether on the spacetime metric, the Trans-Planckian ansatz has a curious prediction for the numerical value of $p_0$, i.e. aether pressure far from the black hole. Comparing the scale of $p_0$ with the density ($\simeq$ $-$ pressure) of the cosmological dark energy, $\rho_{\Lambda}$:
\beq
\frac{p_0}{\rho_{\Lambda}} = -\frac{2}{3}\theta_{P}^{-1} \left(m \over 85 ~\msun \right)^{-3},
\eeq
where we assumed $\Omega_{\Lambda} = 0.7$ and $H_0 = 70 ~{\rm km/s/Mpc}$. \textcolor{black}{The resulting deviation from the Schwarzschild metric is shown in Fig. (\ref{fig-dphi}) for stellar mass black holes.}

This leads us to a very interesting possibility, which was first conjectured in \cite{Afshordi:2008a}: that the formation of stellar-mass black holes could trigger the onset of cosmic acceleration, especially since aether and dark energy have similar pressures, assuming that the aether pressure is set by the Trans-Planckian ansatz for stellar mass black holes. To see this, we can explicitly compare the black hole spacetime (Eqs. \ref{e2} and \ref{e10}) far from the black hole ($r \gg m$):
 \beq
 ds^2 = -(1+2\pi p_0 r^2)^2 dt^2 + dr^2 + r^2 d\Omega^2,
 \eeq
 with the de-Sitter spacetime:
 \beq
 ds^2 = -(1-8\pi \rho_{\Lambda}r^2/3) dt^2 + (1-8\pi \rho_{\Lambda}r^2/3)^{-1} dr^2 + r^2 d\Omega^2.
 \eeq
 We thus notice that non-relativistic particles close to the origin, \textcolor{black}{but far from the black hole horizon} ($ 2m \ll r \ll |p_0|^{-1/2}$) see the same Newtonian potential (or $g_{tt}$) in both spacetimes, if $p_0 = -2\rho_{\Lambda}/3$. In other words, close-by non-relativistic test particles (such as galaxies, stars, or other black holes) accelerate away from the origin/black hole, similar to a de-Sitter space. Moreover, this acceleration will correspond to the current cosmological observations, if the mass of the black holes is roughly:
 \beq
 m \simeq 85 ~\theta_P^{-1/3} ~\msun .\label{mbh_msun}
\eeq

So far, our solution has neglected the effects of black hole spin. Indeed, spin is
expected in realistic black holes, which are fed by astrophysical
accretion disks. For example, the dimensionless spin parameter, $a_* =
a/m$ was recently measured for two stellar-mass black holes, to be
within the range $0.65-0.85$ \cite{2006ApJ...636L.113S}. In order to
include this effect, we conjecture that $p_0$ scales as $T_H^3$ (as suggested in \cite{Afshordi:2008a}), for general black hole spins. This is justified, as the Trans-Planckian ansatz is controlled by the Hawking temperature, $T_H$, while $f(r)$ also depends on the surface gravity close to the black hole horizon, which is also proportional to $T_H$. With this assumption, the scale-dependence should go as:
\beq
\frac{p_0(m,a_*)}{p_0(m,0)} = 8\left[1+\left(1-a^2_*\right)^{-1/2}\right]^{-3},
\eeq
which is in the range $0.2-0.6$ for $a_* = 0.65-0.85$.

\textcolor{black}{While this paper only deals with static vacuum solutions, it was shown in \cite{Afshordi:2008a} that for non-relativistic fluids (e.g. stars, planets): $p' \approx -T^{\alpha}_{\alpha}/4+$const., i.e. the local matter density sets the aether pressure up to a constant. One expects that the constant term would be set by the boundary conditions at infinity, or by cosmology. Alternatively, what we suggest in this section is that the boundary condition can be set at the horizons of the black holes. The fact that this can naturally explain the onset of cosmic acceleration is certainly very suggestive, but the best way to test this hypothesis is to see how/if this boundary condition can emerge from the process of (classical or quantum) gravitational collapse into a black hole. We leave this question to future studies. }

A further implication of this hypothesis is that solar/stellar mass is the minimum mass of black holes allowed in the model. A discovery of significantly sub-solar mass black holes (e.g. primordial black holes with $M \ll \msun$) could potentially rule out the Trans-Planckian ansatz, as it would imply much larger than observed cosmic acceleration rates for $\theta_P \sim 1$.

Of course, we also need to patch together and coarse-grain individual black hole spacetimes into a de-Sitter space, in order to rigorously prove this correspondence. However, the above argument is already very suggestive, as long as there are many black holes within the cosmological/de-Sitter horizon, so that one can trust the above Newtonian argument. In the next section, we provide an approximation to the cosmological spacetime of multiple black holes.

\section{Global Contribution of Multiple Black Holes}
\label{sec:Global}

In this section, we will seek an approach to approximately find the spacetime of multiple black holes with gravitational aether, which can be used to describe an approximate FRW cosmology. Here, for simplicity, we focus on the quasi-static Newtonian regime, where we could assume hydrostatic equilibrium for aether in the vacuum (\ref{hydro}). For simplicity, we ignore the matter in-between black holes \footnote{\textcolor{black}{This is not a bad approximation since, as we argued in the last section, the effect of matter on the aether pressure is localized and does not extend into vacuum in the non-relativistic regime.}}, and assume that black holes are much farther apart than their horizon sizes, but are much closer than the cosmological horizon. In this limit, using Eq. (\ref{hydro}) we have:
\beq
\nabla^2 \ln p = - \nabla^2 \phi = 0,\label{laplace}
\eeq
where the assumption of $\nabla^2\phi =0$ is the equivalent of the Poisson equation in Newtonian gravity, for zero matter density (which also applies to aether). We thus see that fixing the aether pressure in the vicinity of black holes, through the Trans-Planckian ansatz (\ref{p0_m_eq}), is equivalent to solving the Laplace equation (\ref{laplace}) with Dirichlet boundary conditions at (or close) to the horizon of the black holes \footnote{Since pressure approaches $p_0$ at several BH horizon radii for individual black holes, as long as the distance in-between black holes is much larger than their horizon radii, the exact radius at which the boundary condition is set is not important.}.

 This problem is analogous to finding the electrostatic potential between multiple conducting spheres, which can be solved using the Green's function for the appropriate geometry. For a single sphere of radius at the origin (and in a flat space), there is an exact expression for the Green's function, which can be found using the {\it method of images} (e.g. \cite{JAC98a}):
\begin{equation}
G_{D}(\mathbf{x},\mathbf{x'}) = \frac{1}{|\mathbf{x}-\mathbf{x'}|}-\frac{a}{x'|\mathbf{x}-\frac{a^{2}}{x'^{2}}\mathbf{x'}|}.
\label{e12}
\end{equation}
For $n$ spheres (black holes) at positions $\mathbf{x}_{i}$ and with radii $a_i$ ($=2m_i$), we may expand this Green's function, up to first image, as
\bea
&&G_{D}(\mathbf{x},\mathbf{x'}) =  \frac{1}{|\mathbf{x}-\mathbf{x'}|}\nonumber\\&-&\sum_{i=1}^n \frac{a_{i}}{|\mathbf{x'}-\mathbf{x}_i||\mathbf{x}-\mathbf{x}_i-\frac{a_{i}^{2}}{|\mathbf{x'}-\mathbf{x}_i|^{2}}(\mathbf{x'}-\mathbf{x}_i)|} + {\cal O}\left[a^2\over|\mathbf{\Delta x}|^{3}\right],\nonumber\\
\label{e13}
\eea
which is a good approximation, as long as the distance between the spheres/black holes is much larger than their sizes. Now, using Green's theorem, we can find aether pressure in-between the black holes, in terms of the pressure on the surfaces of the spheres, $p_i$'s:
\beq
\ln p(\mathbf{x}) - \ln \bar{p}= -\frac{1}{4\pi}\sum_{i=1}^n\oint_{S_i}  \mathbf{ds}'\cdot\frac{\partial G_D}{\partial \mathbf{x'}} [\ln p_i(\mathbf{x'})- \ln \bar{p}],
\label{e14}
\eeq
where $\ln\bar{p}$ is the log of pressure at infinity, and $\oint_{S_i} \mathbf{ds'}$ are surface integrals over the horizons of the black holes (assuming a flat geometry), while $p_i \propto m^{-3}_i$ are fixed by the masses of the blacks holes, using the Trans-Planckian ansatz (\ref{p0_m_eq}). Since the Green's function (\ref{e13}) is analogous to superposition of electrostatic potentials of point charges, we can use Gauss's theorem to evaluate the surface integrals:
\beq
\ln p(\mathbf{x}) - \ln \bar{p}= \sum_{i=1}^n \frac{a_{i}[\ln p_i(\mathbf{x'})- \ln \bar{p}]}{|\mathbf{x}-\mathbf{x}_{i}|}.
\label{lnp}
\eeq
Now, using the assumption of statistical homogeneity, we expect the spatial/ensemble average of $\ln p$ to be the same as $\ln \bar{p}$. If we take ensemble averages of both sides of Eq. (\ref{lnp}), this yields:
\beq
\ln \bar{p} = \frac{\langle a_i \ln p_i \rangle}{\langle a_i \rangle},
\eeq
or alternatively:
\beq
\bar{p} = -\frac{2}{3} \rho_{\rm DE, eff} = -\frac{1}{256\pi^2 \theta_{P} m^3_*}, ~\ln m_* \equiv \frac{\langle m_i \ln m_i \rangle}{\langle m_i \rangle},
\eeq
where we used $p_i \propto m^{-3}_i$ and $a_i = 2m_i$, as well as Eq. (\ref{p0_m_eq}). In other words, in the presence of multiple black holes, the mean aether pressure, and thus FRW cosmology, is set by $m_*$, which is the mass-weighted geometric mean of black hole masses. Subsequently, the correspondence of this mean aether pressure with an effective {\it Dark Energy} or cosmological constant density was demonstrated in the last section.

Furthermore, taking the Laplacian of Eq.(\ref{lnp}), we can find an equation for the perturbation of effective Dark Energy, for sub-horizon perturbations (but on scales larger than the size of the blacks holes):
\beq
\nabla^2 \delta_{\rm DE,eff} = -8\pi \rho_{\rm BH} \ln(p_i/\bar{p}) ,\label{deltade}
\eeq
where $\delta_{\rm DE,eff}$ is the overdensity of the effective dark energy, while $\rho_{\rm BH}$ is the black hole density. Eq. (\ref{deltade}) can be, in principle, used to track cosmological structure formation and the impact on CMB anisotropies (through the Integrated Sachs-Wolfe effect), but we postpone a study of these effects to future work.

In the next section, we will provide a quantitative picture of how the cosmic history of accretion into stellar and super-massive black holes (or active galactic nuclei) leads to an estimate of $m_*$ as a function of redshift, and its implications for the effective dark energy scenarios.

\section{Cosmic History of Black Holes and Cosmic Acceleration}\label{sec::history}

An up-to-date inventory of cosmic energy at the
present day, including the contribution from stellar-mass and
supermassive black holes, is provided by \cite{FP04}.  In order to measure $m_*(z)$ we need to take
this a step farther and understand the mass distribution of such black
holes, and their redshift evolution.

The mass distribution of stellar--mass black holes is not
well-determined observationally, but estimates are that it is fairly
broad, with a mean of around $\sim 7M_\odot$ \cite{BJCO,PC03}.  We
base our calculations on the theoretical predictions of \cite{FK01},
which show that the distribution can be approximately represented by a
power-law such that the number density of black holes decreases by
a factor 5 between $M=3M_\odot$ and $M=15M_\odot$.  Assuming this
distribution, the average black hole mass is $8.2M_\odot$.  We will
treat the uncertainty in this distribution by varying the slope
sufficiently to alter this mean mass by $\pm 1M_\odot$.
To determine the
redshift evolution in black hole abundance, we use observations of the
cosmic star formation history, from \cite{HB06}.  There is significant
uncertainty in the shape of this history; however it must obey the
integral constraint that the total stellar mass density today be
$\rho_\ast= 0.0027\rho_{\rm crit} = 3.67\times
10^{8}M_\odot$Mpc$^{-3}$, which is known to a precision of $\sim 30$
per cent \cite{FP04}.
We will therefore normalize the black hole number density at $z=0$ to
$1.46\times10^6$ Mpc$^{-3}$ at $z=0$ \cite{FP04}.  We assume that
changes to the initial mass function do not significantly alter the
shape of the star formation rate density evolution, but primarily
affect the number of black holes formed.
By default we assume
a Kroupa IMF \cite{Kroupa}, which is the ``second
model'' considered in \cite{FP04}.  For this choice, 0.19 per cent of
stars formed end up as black holes; a more useful number is that for every solar mass of stars
formed 0.0025 black holes are created.  These numbers change by less
than 5 per cent if we assume a Chabrier IMF \cite{Chabrier}; we expect
therefore the uncertainty on
the normalization of the black hole mass function to be dominated by
the 30 per cent uncertainty in the present day stellar mass function.
Note, however, that a pure Salpeter IMF
\cite{Salpeter} would produce significantly fewer black holes,
only 0.0013 for every solar mass formed.

We base our estimate of the supermassive black hole mass distribution
on observations of the quasar luminosity function.  This requires
assumptions about the lifetime and obscuring column density of quasars;
for this we adopt the model of \cite{PH+06} who describe a
merger-driven scenario of black hole growth.  Using this model, the
$z=0$ mass density of supermassive black holes is
$2.9^{+2.3}_{-1.2}\times10^5M_\odot$Mpc$^{-3}$.  This is somewhat
smaller than the value of $5.4\times10^5M_\odot$Mpc$^{-3}$
determined from the correlation between black hole mass and bulge
luminosity \cite{F02,Gebhardt}, as computed by \cite{FP04}.  However,
the uncertainty on the latter is a factor of two, and a lower value of
$3.4\times10^5M_\odot$Mpc$^{-3}$ is obtained \cite{FP04} if one uses the correlation with
velocity dispersion for early type galaxies \cite{MF01,T02} rather
than luminosity.

\begin{figure}
\includegraphics[width=\linewidth]{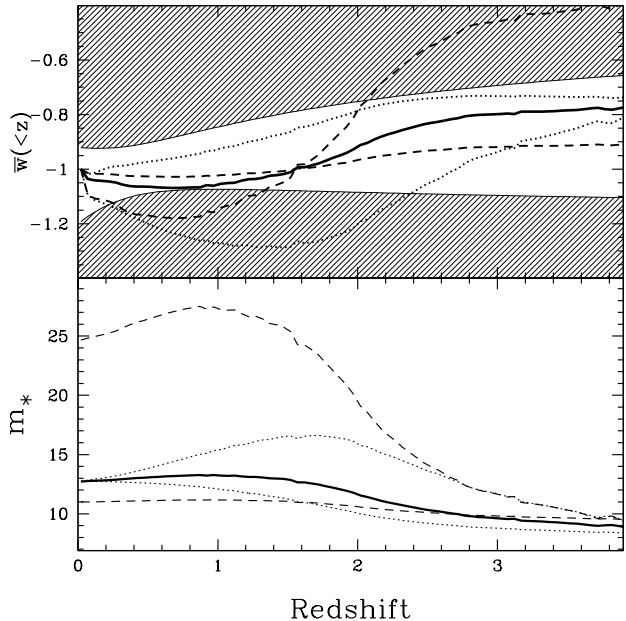}
 \caption
{
{\bf Bottom panel:} The mass-weighted geometric mean of black hole
masses, $m_*$, in units of $\msun$ as a function of redshift.  Our fiducial model (solid, black
line) assumes our best estimates of the mass distribution evolution of
the black hole mass distribution.  Dashed lines indicate the range of
uncertainty expected due to the unknown relative contribution of
supermassive and stellar-mass black holes at $z=0$, while the dotted
lines represent the uncertainty in the shape of the star formation
density evolution from \cite{HB06}. {\bf Top panel:} The prediction of
the equation of state parameter $\bar{w}(<z)$ from Equation~\ref{eqn-w}, for
the same models.  The dashed area shows the region excluded at 68\% confidence level for this
parameter, as measured from independent observations \cite{2009ApJS..180..330K}.
}\label{fig-obsmstar}
\end{figure}

With this in hand we are able to compute the expected $m_*(z)$, and
this is shown in the bottom panel of Figure~\ref{fig-obsmstar}.  Our
best estimate of the local, mass-weighted geometric mean of black hole
masses is $m_*(0)= 12.7M_\odot$.  The dashed lines represent the
range of uncertainty on this $z=0$ normalization.  A larger value of
$m_*$ is obtained by reducing the contribution of stellar-mass black
holes (assuming the local density is 30\% lower than our fiducial
model, and assuming the mass distribution is more steeply weighted to
lower masses, so the average mass is $7.2M_\odot$), and increasing the
contribution of supermassive black holes (by increasing the $z=0$
space density within the $1\sigma$ uncertainty, to
$5.2\times10^5M_\odot$Mpc$^{-3}$).  This yields $m_*(0)= 24.7M_\odot$.
Pushing the numbers in the opposite direction, we obtain $m_*(0)=
10.5M_\odot$. Using Eq. (\ref{mbh_msun}) for the current effective
density of  dark energy, and ignoring the spin of black holes, this
range in $m_*(0)$ translates to a range for the Trans-Planckian parameter $\theta_P$:
\beq
\theta_P = (0.4-5) \times 10^2.\label{theta_p}
\eeq
We can consider spinning black holes, using our scaling argument from
Section \ref{trans-planckian} and taking a nominal value of $a_*=0.75$.  This implies a lower
range for the Trans-Planckian parameter, $\theta_P = 20-300$, in order to
match the current rate of cosmic acceleration. The fact that $\theta_P
\sim 1$, further justifies a Trans-Planckian, or quantum gravitational
origin for the observed ``dark energy phenomenon''.

The evolution of the stellar-mass black hole mass density is dependent
upon the shape of the star-formation-rate density plot from
\cite{HB06}.  To consider the effect of this, we construct two star
formation histories that are consistent with those data within the
1$\sigma$ error bars, but which produce as many stars as possible at
either high redshift $(z>1)$ or at low redshift ($z<0$).  We still
renormalize this to match the local stellar mass density.  These
extremes are shown in Figure~\ref{fig-obsmstar} as dashed lines.  The
evolution of the supermassive black hole distribution is very model
dependent, and not well constrained.  We note that the two different
predictions shown by \cite{PH+06}, which make different assumptions
about the quasar space density evolution at $z>2$, have a subdominant
effect on the predictions shown here, relative to the other
uncertainties considered.

Within an effective dark energy description of FRW cosmology, a fixed
dark energy equation of state, $w$, implies that dark energy density
evolves as $(1+z)^{3(1+w)}$, as a function of redshift, $z$. The
effective equation of state (which is simply a way to parameterize
cosmic expansion history) is observationally constrained to
\beq
w(z) = -1.06 \pm 0.14 + (0.36 \pm 0.62)\frac{z}{1+z},\label{w(z)}
\eeq
at 68\% confidence level, based on cosmic microwave background,
baryonic acoustic oscillations, and supernovae Ia observations,
assuming a spatially flat cosmology \cite{2009ApJS..180..330K}, and a
linear dependence of $w(z)$ on the cosmological scale factor
$=(1+z)^{-1}$.

We can define a mean equation of state as:
\bea\label{eqn-w}
1+{\bar w}(<z) \equiv \textcolor{black}{\frac{1}{3}\frac{\ln[\rho_{\rm DE, eff}(z)/\rho_{\rm DE, eff}(0)]}{\ln(1+z)} }\nonumber\\= -\frac{\ln[m_*(z)/m_*(0)]}{\ln(1+z)},
\eea
since,  $\rho_{\rm DE, eff}(z) \propto m^{-3}_*(z)$, as we saw in the
last section. We show this estimate of $w$ for the models described
above, in the top panel of Figure~\ref{fig-obsmstar}.  Our fiducial
model predicts a value of $w$ that deviates from $-1$ by less than 5
per cent out to $z\sim 2$, but predicts it should reach $w=-0.8$ by
$z=3$.  There is considerable uncertainty on this, however, due both
the unknown distribution of black hole masses at $z=0$ (dashed lines)
and the unknown shape of the star formation rate density evolution
(dotted lines).

While most these models are consistent with the current bounds on the effective dark energy equation of state (using Eq. \ref{w(z)}):
\beq
{\bar w}(< z) = -1.06 \pm 0.14 + (0.36 \pm 0.62)\left[1-\frac{z}{(1+z)\ln(1+z)}\right],
\eeq
stage IV dark energy missions, as quantified by the dark energy task force report \cite{2006astro.ph..9591A}, are expected to have percent level sensitivity to ${\bar w} (< 1-3)$, and thus should be able to distinguish the aether model with these $m_*(z)$ histories from a cosmological constant.

\section{Conclusions and Future Prospects}
\label{sec:conclusions}

We have shown that static black hole solutions exist in the
gravitational aether model of \cite{Afshordi:2008a}. The model is an
attractive alternative to the cosmological constant, which does not
suffer from the tremendous fine-tuning problem of vacuum energy in
standard model. We find that in the presence of a gravitational aether,
the Schwarzschild black hole is sufficiently perturbed so as to result
in a Trans-Planckian connection between physics near the black hole
horizon and cosmology. This could be a phenomenological product of
quantum gravity, and it naturally explains the present-day acceleration of
cosmic expansion as a result of formation of stellar/solar-mass black
holes.

Indeed, the recent discovery of cosmic acceleration, or dark energy \cite{Riess:1998cb, Perlmutter:1998np} might be the first concrete evidence for quantum gravity and/or Trans-Planckian physics. Future work may include an exploration of quantum properties of this black hole solution. In particular, a natural next step would be to understand how quantum gravity can resolve the null singularity at the event horizon.

As discussed in Sec. \ref{trans-planckian}, another important question yet to be addressed is whether dynamical evolution could lead to the static solutions found in this work.   
While prior to formation of black holes, the integration constant $p_0$ is set by large-scale conditions, as black hole horizons form, we speculate that the constant is instead set by conditions at the event horizon. In order to understand the causal transition between these two boundaries, and how fast the effect will propagate away from the black hole, a more complete dynamical picture is necessary. 

Furthermore, in the presence of multiple black holes with relative motion, the aether is expected to be locally dragged by different black hole horizons. However, for black holes at large separations compared to their horizon sizes and non-relativistic velocities (as expected in astrophysical situations), the perturbations to the static solution is expected to be small. 

To conclude, we would like to entertain the exciting possibility that the gravitational aether \cite{Afshordi:2008a} might provide a complete solution to the three aspects of the cosmological constant (CC) problem, as discussed in the Introduction:
\begin{enumerate}
 \item  {\it Old CC problem}: Gravitational aether theory decouples quantum vacuum from geometry, which allows a nearly flat spacetime even in the presence of large vacuum energy densities expected from the standard model of particle physics. The model makes specific predictions for physics at big bang nucleosynthesis and radiation-matter transition era, which will be tested with precision cosmological probes over the next decade \cite{Afshordi:2008a}.

\item {\it New CC problem}: Formation of black holes leads to a UV-IR coupling, which connects near-horizon Planck-scale physics to cosmology, and can naturally lead to cosmic acceleration, even in the absence of a real dark energy component.

\item {\it Coincidence problem}: As we showed in Sec. (\ref{sec::history}), the stellar mass black holes expected in standard star formation, can naturally lead to the observed present-day acceleration of the Universe. The competition between the contribution of stellar mass black holes, and super-massive black holes leads to an evolution of the effective dark energy density, which can be tested with NASA's future {\it Joint Dark Energy Mission (JDEM)} \footnote{\href{http://jdem.gsfc.nasa.gov/}{http://jdem.gsfc.nasa.gov/}} or its European counterpart {\it Euclid} \footnote{\href{http://sci.esa.int/science-e/www/object/index.cfm?fobjectid=42266}{http://sci.esa.int/science-e/www/object/index.cfm?fobjectid=42266}}.

\end{enumerate}

We would like to acknowledge helpful discussions with Simon DeDeo, Ghazal Geshnizjani, Jurjen Koksma, Rob Myers, Lee Smolin, and Mark Wyman.
CP and NA are supported by Perimeter Institute for Theoretical Physics.
 Research at Perimeter Institute
is supported by the Government of Canada through Industry Canada and by
 the Province of Ontario through
the Ministry of Research \& Innovation.

\bibliographystyle{utcaps_na2_ads}
\bibliography{aether_BH}

\end{document}